# Automated Test Production Systematic Literature Review

Gomes, J.M.[1] and Dias, L.A.V.[1]

[1]Instituto Tecnológico de Aeronáutica - ITA

January 5, 2024

**Abstract**

Identifying the main contributions related to the *Automated Test Production* of Computer Programs and providing an overview about models, methodologies and tools used for this purpose is the aim of this *Systematic Literature Review*. The results will enable a comprehensive analysis and insight to evaluate their applicability. A previously produced *SLM (Systematic Literature Mapping)* contributed to the formulation of the "Research Questions" and parameters for the definition of the qualitative analysis protocol of this review.

## 1 Objectives

The broader goal of this research, while on the one hand is to obtain the State of the Art in *ATP (Automated Test Production)*, find the problems faced and track the progress of researchers in the field, on the other hand we also intend to list and categorize the *ATP* methods, techniques and tools that meet the needs of professionals producing specialized business applications for internal use within their corporations - eventually extending to the needs of professionals in companies specializing in the production of generic or even academic computer applications.

Providing the scientific and technological community with an overview of models, methodologies and tools used for this purpose is the goal of this *SLR (Systematic Literature Review)*. The results will allow a comprehensive analysis and insight to evaluate their applicability.

## 2 Systematic Literature Review

### 2.1 Planning

In order to analyze, evaluate and interpret the available research for the matter of *ATP* of Computer Programs, we applied the method proposed by Brereton *et*

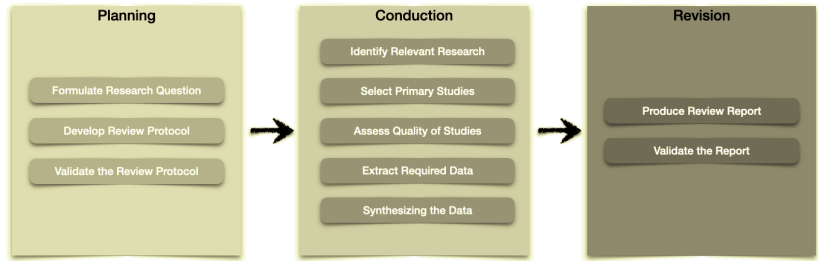

Figure 1: Steps to perform an *SLR* (adaptedo from [1])

*al.* and presented in the Figure 1. Whereas the *Systematic Literature Mapping*[2] survey gave us an overview and insight into what represents current research in *ATP*, with the present *SLR* we aim to identify the effort needed to answer our research questions[1].

Several adjustments suggested by Kitchenham *et al.* led us to consider the methodologies of Basili and Weiss for formulating the research questions, of Chen and Babar for classifying the type of study, of Keele *et al.* for conducting a quality assessment of each proposal, and of Kitchenham *et al.*, Dyba *et al.* for gauging the scientific rigor of the studies obtained and in accordance with criteria of interest to our research [3, 4, 5, 6, 7, 8].

We defined the questions that this research intends to answer by following the Basili and Weiss Basili and Weiss - a systematic method for organizing measurements. The method begins by specifying an objective (purpose, object, problem, point of view). The goal is refined into several questions, each of which in turn is refined into Basili and Weiss metrics. Providing the answers to the questions, the data can be analyzed to identify whether or not the objectives were met [4].

Thus, the goal of this *SLR* is:

- **Purpose** *Understand and characterize ...*
    - **Problem** *... the solutions in* ATP *...*
        * **Object** *... of Computer Program Testing ...*
            · **Point of view** *... used by researchers and informatics professionals.*

Based on the above stated goal, we derive the questions to be answered (see Table 1).

| # | Question |
|---|---|
| **QP1** | What is the model, process, framework or tool used for test production? |
| **QP1.1** | What type of test is generated? |
| **QP1.2** | What test production techniques were applied? |
| **QP1.3** | What tools does the study employ? |
| **QP1.4** | What are the prerequisites for its application? |
| **QP1.5** | What types of studies or evaluations have been conducted? |

Table 1: Research questions from *SLR*

The question **QP1** allows us to categorize the approaches and learn about the applicability of the study proposal and meets the requirement of identifying solutions to the problem.

| Name |
|---|
| *IEEE Xplore* |
| *ACM Digital Library* |
| *Google Scholar* |
| *CiteSeerX* |
| *Inspec* |
| *ScienceDirect* |
| *EI Compendex* |
| *Springer Link* |

Table 2: Scientific publication bases

## 2.2 Conduction

The studies obtained for conducting this *SLR* were obtained from the scientific publication sources listed in the Table 2 and selected in our *SLM*[2].

The selection of documents was based on Inclusion and Exclusion Criteria defined iteratively during the reading of the documents found and provided to determine the suitability of each to the objectives of this work. The Inclusion Criteria are those presented in the Table 3, and in the Table 4 we have the Exclusion Criteria.

| # | Descrição |
|---|---|
| **CI1** | The Study is a review of current models/processes/frameworks |
| **CI2** | The study is useful for a better understanding of the problem |
| **CI3** | The study discusses one or more tools within a process applicable to this research |
| **CI4** | The study proposes a model/process/framework/methodology |

Table 3: Inclusion criteria for the *SLR*

| # | Descrição |
|---|---|
| **CE1** | The study presents some model/process/framework/methodology but does not provide enough information about its use |
| **CE2** | The study does not contain any kind of evaluation for the demonstration of results, such as case studies, experiments, or examples of use |
| **CE3** | The study is not directly related to this research |

Table 4: Exclusion criteria from *SLR*

Given the large number of studies[2] found we organized our work into iterative steps:

1. Reading the summary and conclusion; and

2. Selection by reading the entire document.

| Name | Qty. | % |
|---|---|---|
| **IEEE Xplore** | 33 | 14.73 |
| **ACM Digital Library** | 66 | 29.46 |
| **Springer Link** | 11 | 4.91 |
| **Periódicos da CAPES** | 114 | 50.89 |
| Sub-total | 224 | |
| *Duplicate studies* | 7 | 4.24 |
| *Rejected studies* | 52 | 31.52 |
| **TOTAL** | 165 | |

Table 5: Result of the selection of primary studies for the *SLR*

After reading the abstract of the pre-selected publications, if they are interesting, we also read the conclusion, and if the paper remains promising to our expectation it is then selected (see Figure 1 and Table 5[2]). The selection criteria at this stage become slightly stricter and we are now looking for model-oriented papers, methods, techniques and tools that generate tests for general-purpose computer applications and we avoid for example:

- Texts that deal with test generation but never in an automated or semi-automated way;

- Tests applied to computing devices or equipment of specific use such as those of industrial, embedded and specialized use such as automotive and so on;

- Tests related to a specific or particular application such as a database manager or a commercial application, etc;

- Tests aimed at a specific use of a technology or facility, such as geo-location or the use of gestures on touch screens, etc;

- Testing targeted at a particular industry or segment of the economy or society, such as healthcare, banking and finance, trade, etc.; and

- Tests oriented to the detection of non-functional flaws, such as security, performance, etc.

Of course, despite the rigor we seek to conduct this research, exceptions will be made for the sake of completeness and common sense, and we may include studies, which applied to the technologies or segments listed above, are part of a case study within a larger context.

A difficult decision to make at this point is whether or not to discriminate the application of tests on a given platform. Nowadays we see the proliferation of the use of computing resources in a pervasive way, and we have computing devices on our desks, in our pockets, like jewelry on our wrists, and even in our home appliances [9]. And this is just one of the aspects we observe directly, since the computing itself can take place inside the device locally, or remotely on servers connected to the Internet, or even in hybrid mode. Following the criteria listed above, we will discriminate particular or niche or proprietary uses of certain manufacturers and not consider them, but we must include the cases where these platforms are extremely widespread and have a wide range of applications developed on them, such as those used in cell phone handsets for example.

After this preliminary selection we will proceed with a more careful reading of the complete document that will generate notes that will be useful in this work.

In the full reading of the documents obtained after the first and second search criteria, the selection becomes a qualitative analysis where each document is evaluated according to quality criteria and scored, a method proposed by Keele *et al.* and which we used to evaluate quality of studies for this review [6]. These new criteria, which we call Quality Criteria, are based on the research questions listed in the Table 1 and whose answers will be mapped into points. The sum of the points obtained by each document will give us a score, within which we define a cut-off value that will select or eliminate the document.

The questions defined for this selection phase are those listed in the Table 6.

| # | Description | Questions |
|---|---|---|
| **CQ1** | Describe the model, process, framework or tool used or created? | |
| **CQ1.1** | Does it show the types of tests generated? | **QP1.1** |
| **CQ1.2** | Does it present the test generation techniques used? | **QP1.2** |
| **CQ2** | List the prerequisites? | **QP1.3, QP1.4** |
| **CQ2.1** | Does it list technologies and knowledge required for proper use? | |
| **CQ2.2** | List situations for which its use is recommended? | |
| **CQ2.3** | List situations for which the use is not recommended? | |

Table 6: Quality criteria of the *SLR*

The possible answers to the questions presented in the Table 6 are listed in the Table 7. With the quality assessment we answered most of the research questions proposed in the Table 1 with the exception of the question "**QP1.4** - *What types of studies or evaluations have been conducted*", for which we adopted the classification into six categories proposed by Chen and Babar and listed in the Table 9 [5].

| Answer | Value |
|---|---|
| Yes | 1.0 |
| Partially | 0.5 |
| No | 0.0 |

Table 7: Quality criteria values

Based on the questions in the Table 6 and the answer values listed in the Table 7 we have a **Max Score** of **10 points** (only elementary questions count towards the maximum score). We set the cutoff score at **cutoffScore points**, i.e. documents that do not get at least this score in the Qualitative Analysis phase will be discarded.

| Criteria | Weight |
|---|---|
| CQ1 | 7 |
| CQ2 | 3 |

Table 8: Quality criteria weights

We apply the weights listed in the "Table 8 and thus a description of the requirements in the studies ("**CQ1**") of the models or tools presented have higher objective value than the description ("**CQ2**") or the listing of their prerequisites.

| # | Description |
|---|---|
| **RA** | *Rigorous Analysis* |
| | Rigorous derivation and proof suitable for formal model |
| **CS** | *Case Study* |
| | An empirical investigation of a current phenomenon in its actual context; when the boundaries between phenomenon and context are not clearly evident; and in which multiple sources of evidence are used |
| **DC** | *Discussion* |
| | Qualitative and textual opinion |
| **EX** | *Example* |
| | The authors describe an application and provide an example to help in the description, but the example is "used to validate" or "evaluate" to the extent that the authors suggest |
| **ER** | *Experience Report* |
| | The result has been used in real examples, but not in the form of case studies or controlled experiments, evidence of its use is collected informally or formally |
| **FS** | *Field Study* |
| | Controlled experiment conducted in industry settings |
| **LH** | *Laboratory Experiment with Human Subjects* |
| | Identifying precise relationships between variables in a controlled environment using humans and applying quantitative techniques |
| **LS** | *Laboratory Experiment with Software Subjects* |
| | A laboratory experiment to compare the performance of the proposed model or tool with existing ones |
| **SI** | *Simulation* |
| | Running with stochastic data and using a real model or tool |

Table 9: Categories of studies evaluated in the SLR

With the adoption of the categorization of studies that we have listed in Table 9, it only remains to quantify these categories so that we can properly score the studies and select those of interest to us. For this Galster *et al.* suggests an analysis in three dimensions[10].

- **C** - Context, which refers to the environment in which conducted, includes the experience of the personnel involved, processes used, and illustrates the feasibility of applying a particular technology

- **D** - Design, which describes the products, resources and processes used in the study

- **V** - Validity, that is, a discussion of the validity of the results obtained, their limitations, and what threatens this validity

We capture the value of each dimension as proposed in Ivarsson and Gorschek in three levels [11]:

- **1** - weak;

- **2** - regular; and

- **3** - strong.

This rigor is appropriate for empirical studies and meets the ideas presented by Kitchenham *et al.* and elaborated on in Dyba *et al.* Cross-referencing the dimensions with the values we arrive at the Table 10 with the classification of rigor [7, 8].

| Dimension | Value | | |
|---|---|---|---|
| | Weak (1) | Regular (2) | Strong (3) |
| Context of the study (C) | There seems to be no description of the context in which the evaluation is conducted | The context in which the study is conducted is mentioned or briefly presented, but is not described in a way that the reader can understand it and compare it to another context | The context is described in a way that the reader can understand it and compare it to another context |
| Study design (D) | There seems to be no description of the evaluation design presented | The study design is described briefly, e.g., "10 students turned in assignments 1, 2, and 3 on time" | The study design is described in a way that a reader can understand, among other things, the variables measured, the control used, the treatments, the selection and sampling used, and other pertinent information |
| Validity of the discussion of results (V) | There seems to be no description of any threats to the validity of the evaluation | The validity of the study is mentioned, but not described in detail | The validity of the evaluation is discussed in detail where threats are described and measures to limit them are detailed |

Table 10: Classification of rigor applied to studies in the SLR

## 2.3 Qualitative Analysis

Based on the parameters set in the Tables 6, 7 and 10 the selected studies (see Table 5) were qualified and the result can be seen in the "Selected Primary Studies".

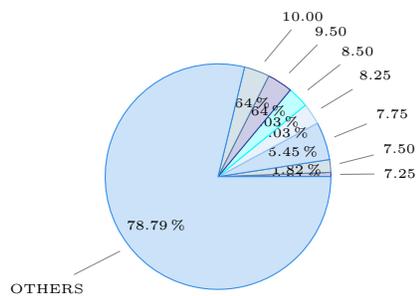

| Points | % | Qty. |
|---|---|---|
| 7.25 | 0.61 | 1 |
| 7.50 | 1.82 | 3 |
| 7.75 | 5.45 | 9 |
| 8.25 | 3.03 | 5 |
| 8.50 | 3.03 | 5 |
| 9.50 | 3.64 | 6 |
| 10.00 | 3.64 | 6 |
| OTHERS | 78.79 | 130 |

Table 11: Qualified studies

Figure 2: Qualified studies

In the Figure 2 and Table 11 we list the totals of studies that have qualified based on the cutoff score (**7 points**).

## 2.4 Analysis of Results

Based on the criteria listed in the Table 3 165 were selected and 52[2] rejected.

The research questions listed in Table 1 were applied to the selected studies and we obtained the results that we list further below.

### 2.4.1 Models, processes, frameworks or tools used for test production (QP1)

In the *SLM*[2] we applied the Petersen *et al.* systematics to classify the obtained documents as seen in the Figure 3 and evaluated the type of contribution based on interpretation of the abstracts and listed in the Figure 4 and Table 17 [12, 13].

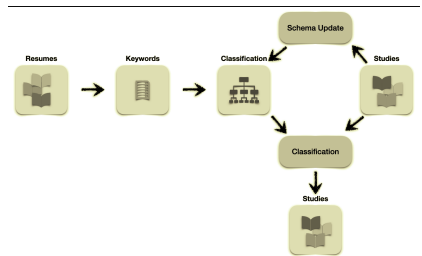

Figure 3: Classification Scheme (adapted from [12])

**What types of tests are produced (QP1.1)?** In the Figure 4 we list the types of generators addressed by the studies, and in the Figure 5 the types of tests produced.

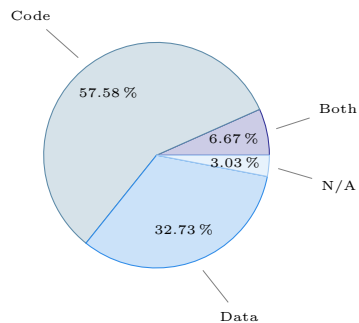

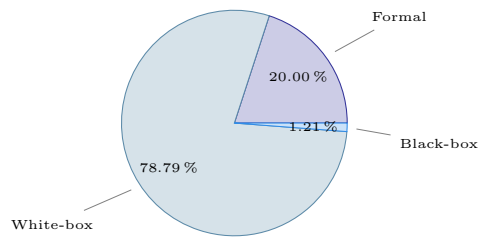

Figure 5: Generators Types

Figure 4: Artifact Types

| Artifact Type | % | Qty. |
|---|---|---|
| Both | 32 | 6.67 |
| Code | 5 | 57.58 |
| Data | 124 | 32.73 |
| N/A | 4 | 3.03 |

Table 12: Types of Artifacts Produced

| Generator Type | % | Qty. |
|---|---|---|
| Formal | 33 | 20.00 |
| White-box | 130 | 78.79 |
| Black-box | 2 | 1.21 |

Table 13: Types of Generators

**What test generation techniques were applied (QP1.2)?** The most used techniques found in the studies can be seen in Figure 6.

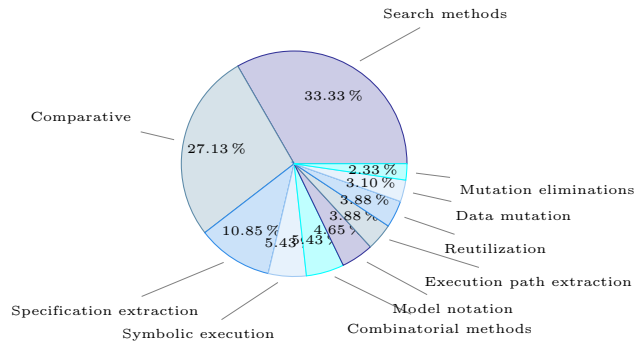

Figure 6: Models, methods, processes and techniques applied in *ATP*

**What tools does the study employ (QP1.3)?** In Figure 7 we list the tools employed by the studies.

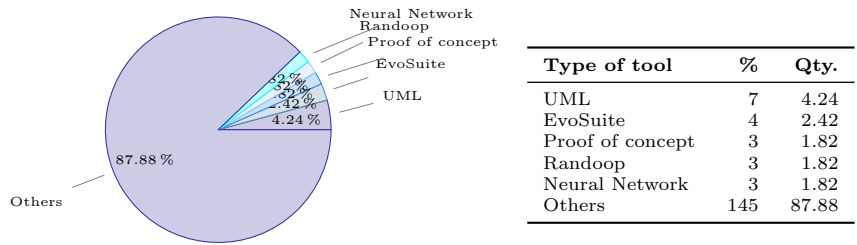

| Type of tool | % | Qty. |
|---|---|---|
| UML | 7 | 4.24 |
| EvoSuite | 4 | 2.42 |
| Proof of concept | 3 | 1.82 |
| Randoop | 3 | 1.82 |
| Neural Network | 3 | 1.82 |
| Others | 145 | 87.88 |

Table 14: Tools used in the studies

Figure 7: Tools used in the studies

We classified as "**Proof of concept**" the studies that used some program to present their approach but did not use or produce a tool that can be considered ready for use by industry and in some even for other studies.

**What are the prerequisites for its application (QP1.4)?** What artificats we observed that are required to perform the analysis according to the study.

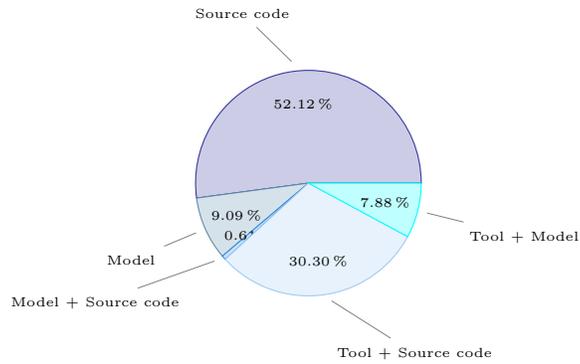

Figure 8: Study Requirements

| Study type | % | Qty. |
|---|---:|---:|
| Source code | 52.12 | 86 |
| Model | 9.09 | 15 |
| Model + Source code | 0.61 | 1 |
| Tool + Source code | 30.30 | 50 |
| Tool + Model | 7.88 | 13 |

Table 15: Study Requirements

**What types of studies or evaluations have been conducted (QP1.5)?**
The types of studies were classified according to Table 16 and the results listed in Figure 9 and Table 17.

| Category | Description |
| --- | --- |
| Metric | A system or standard of measures or measurements taken using an existing standard. |
| Tool | A device or implementation, used to perform a certain function. |
| Model | A comprehensive and systematic approach that includes theoretical principles, benefits and drawbacks, objectives, methodological guidelines and specifications, and the characteristic use of certain sets of methods and techniques. |
| Method | No particular theoretical orientation is inferred in a method. Researchers impose their own particular theoretical beliefs on an experiment when they design and implement it by applying one or more techniques. |
| Technique | A single operation or interaction in which a researcher uses one or more procedures to elicit an immediate reaction from the object of study or to shape the experiment and obtain results. |
| Procedure | An organized sequence of operations and interactions that a researcher uses to conduct an experiment. |
| Intervention | Purposefully interferes with or mitigates various aspects of the object of study and that affect the outcome by applying procedure, technique. The elements acting on the industry during a particular intervention are most often computer applications, the researcher, or both. |
| Approach | A broad way of addressing an industry concern or problem. A specific methods is not implied, but a specific set of techniques will likely come into play when trying to intervene in the industry and the problem that is the subject of the research. The procedures to be used will be determined by the delimitations of the methodological variant in which we design the study. |
| Strategy | An action plan designed to achieve an overall goal. |

Table 16: Type of contribution

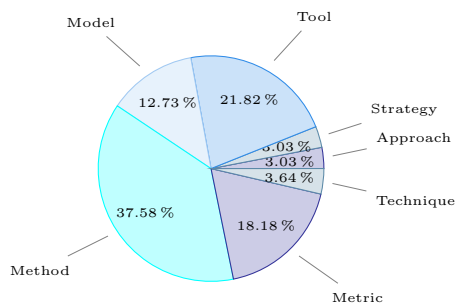

| Tipo Contrib. | Qtd. | Pct. |
| --- | --- | --- |
| Approach | 5 | 3.03 |
| Strategy | 5 | 3.03 |
| Tool | 36 | 21.82 |
| Model | 21 | 12.73 |
| Method | 62 | 37.58 |
| Metric | 30 | 18.18 |
| Technique | 6 | 3.64 |

Table 17: Contributions of the Studies

Figure 9: Contributions of the Studies

# 3 Results

We present the results of a *SLR* employed to find relevant studies on *ATP*. This review applied the methodology of Petersen *et al.* with elements of Basili and Weiss, Keele *et al.*, Brereton *et al.* The results obtained list different models, processes, frameworks and tools used with diverse approaches and results.

## 3.1 Conclusions

Based on the research questions developed in Section 2 we conclude that:

### 3.1.1 Models, processes, frameworks or tools applied to test production

**Types of tests produced**  As we can see from the Figures 4 and 5 much of the studies have been devoted to producing test verification of *white-box*[1]. We also observe significant concern with the Oracle problem (determining the correct outputs for given inputs) and there are a significant number of approaches that produce data for program verification.

Given the aspect of this research and the questions we address, *black-box* tests[2] were not representative in our results, however according to Gaudel formal methods were a highlight and these can be considered to be *black-box* tests.

**Techniques applyed to test production**  Several techniques were employed in test production, and among them we highlight **Search Methods** (see Figure 6). Secondly we note the application of **Specification Extraction**[3] as an important approach employed by the studies.

Highlighted we can also observe: **Symbolic Execution**, **Combinatorial Methods** and **Model Notation**.

**Tools applied by the studies**  In the Figure 7 and Table 14 we observed a great diversity of tools used and none in particular stood out in our study, highlighting the use of *UML (Unified Modelling Language)* by a significant percentage (if compared to other approaches) of studies. This fragmentation demonstrates, on the one hand, a great wealth of available solutions and approaches, but it may also be indicative of a lack of maturity of the solutions presented, which can be observed by the opinion of some authors [15, 16].

**Pre-requisites to the proposed approaches**  For even obvious reasons (related to *white-box* testing - see Figure 5), the vast majority of studies require access to the source code of the applications (see Figure 8 and Table 15). Many studies are conducted using tools (none in particular as we have noted - see Figure 7), but an important number have been conducted from models (previously produced by a development process or *ad hoc*).

---

[1] Method of validating non-functional, internal aspects of a computer application.
[2] Method of validating functional and external aspects of a computer application
[3] Comparative studies do not present particular techniques or methods

**Types of studies or evaluations conducted**  In the Figure 9 and Table 17 we list the main contributions of the studies, and note a rather encouraging number of concrete contributions (in the form of methods, tools, metrics, and models) that can be leveraged and extended by the industry.

### 3.2 Future work

This work aimed to prepare the ground for further research on *ATP* where we will determine the challenges in applying generative testing techniques and evaluate the solutions we intend to address.

## Acronyms

**ATP**        Automated Test Production  - pages: 1, 2, 13, 14

**SLM**        Systematic Literature Mapping  - pages: 1, 3, 9
**SLR**        Systematic Literature Review  - pages: 1–3, 13

**UML**        Unified Modelling Language  - page: 13

# Appendices

## Appendix A  Selected Primary Studies

| # | Title | Author |
|---|---|---|
| 1 | Automated Unit Test Generation for Evolving Software | Shamshiri, Sina |
| 2 | Multifaceted Test Suite Generation Using Primary and Supporting Fitness Functions | Gay, Gregory |
| 3 | Improving Model-Based Test Generation by Model Decomposition | Arcaini, Paolo; Gargantini, Angelo; Riccobene, Elvinia |
| 4 | Automatic Generation of Search-Based Algorithms Applied to the Feature Testing of Software Product Lines | Filho, Helson L. Jakubovski; Lima, Jackson A. Prado; Vergilio, Silvia R. |
| 5 | Fuzzy Adaptive Teaching Learning-Based Optimization Strategy for GUI Functional Test Cases Generation | Din, Fakhrud; Zamli, Kamal Z. |
| 6 | One-Size-Fits-None? Improving Test Generation Using Context-Optimized Fitness Functions | Gay, Gregory |
| 7 | Automated Test Case Generation as a Many-Objective Optimisation Problem with Dynamic Selection of the Targets | Panichella, Annibale; Kifetew, Fitsum Meshesha; Tonella, Paolo |
| 8 | Decomposition-Based Approach for Model-Based Test Generation | Arcaini, Paolo; Gargantini, Angelo; Riccobene, Elvinia |
| 9 | Generating Complete Controllable Test Suites for Distributed Testing | Hierons, Robert M. |
| 10 | Exploiting Model Morphology for Event-Based Testing | Belli, Fevzi; Beyazit, Mutlu |
| 11 | AMOGA: A Static-Dynamic Model Generation Strategy for Mobile Apps Testing | Salihu, Ibrahim-Anka; Ibrahim, Rosziati; Ahmed, Bestoun S.; Zamli, Kamal Z.; Usman, Asmau |
| 12 | Detecting Trivial Mutant Equivalences via Compiler Optimisations | Kintis, Marinos; Papadakis, Mike; Jia, Yue; Malevris, Nicos; Le Traon, Yves; Harman, Mark |
| 13 | Deductive Software Verification: From Pen-and-Paper Proofs to Industrial Tools | Hähnle, Reiner; Huisman, Marieke |
| 14 | A multi-objective test data generation approach for mutation testing of feature models | Matnei Filho, Rui A.; Vergilio, Silvia R. |
| 15 | Automatic generation of valid and invalid test data for string validation routines using web searches and regular expressions | Shahbaz, Muzammil; McMinn, Phil; Stevenson, Mark |
| 16 | Achieving scalable mutation-based generation of whole test suites | Fraser, Gordon; Arcuri, Andrea |
| 17 | COMPARATIVE BENCHMARkING OF CONSTRAINTS T-WAY TEST GENERATION STRATEGY BASED ON LATE ACCEPTANCE HILL CLIMBING ALGORITHM | Zamli, Kamal Z.; Abdul Rahman Alsewari; Basem Al-Kazemi |
| 18 | An experimental study of hyper-heuristic selection and acceptance mechanism for combinatorial t-way test suite generation | Zamli, Kamal Z.; Din, Fakhrud; Kendall, Graham; Ahmed, Bestoun S. |
| 19 | MCCFG: an MOF-based multiple condition control flow graph for automatic test case generation | Son, Hyun; Park, Young; Kim, R. |
| 20 | A Tabu Search hyper-heuristic strategy for t-way test suite generation | Zamli, Kamal Z.; Alkazemi, Basem Y; Kendall, Graham |
| 21 | Petri net based test case generation for evolved specification | Ding, Zuohua; Jiang, Mingyue; Chen, Haibo; Jin, Zhi; Zhou, Mengchu |
| 22 | Sentinel: A Hyper-Heuristic for the Generation of Mutant Reduction Strategies | Guizzo, Giovani; Sarro, Federica; Krinke, Jens; Vergilio, Silvia R. |
| 23 | Automated generation of (F)LTL oracles for testing and debugging | Pill, Ingo; Wotawa, Franz |
| 24 | How to Optimize the Use of SAT and SMT Solvers for Test Generation of Boolean Expressions | Arcaini, Paolo; Gargantini, Angelo; Riccobene, Elvinia |
| 25 | Choosing the fitness function for the job: Automated generation of test suites that detect real faults | Salahirad, Alireza; Almulla, Hussein; Gay, Gregory |
| 26 | MoFQA: A TDD Process and Tool for Automatic Test Case Generation from MDD Models | Linda Riquelme; Magalí González; Nathalie Aquino; Luca Cernuzzi |
| 27 | Automatic test cases generation from business process models | Yazdani Seqerloo, Arezoo; Amiri, Mohammad; Parsa, Saeed; Koupaee, Mahnaz |
| 28 | Test Case Generation for Boolean Expressions by Cell Covering | Lian Yu; Wei-Tek Tsai |
| 29 | Transition coverage based test case generation from state chart diagram | Pradhan, S.; Ray, Mitrabinda; Swain, S.K. |
| 30 | A Memetic Algorithm for whole test suite generation | Fraser, Gordon; Arcuri, Andrea; McMinn, Phil |
| 31 | Pairwise test data generation based on flower pollination algorithm | Nasser, Abdullah B.; Alsewari, A.A.; Tairan, N.M.; Zamli, Kamal Z. |
| 32 | Verification and code generation for invariant diagrams in Isabelle | Preoteasa, V; Back, Rj; Eriksson, J |
| 33 | Tri-modal under-approximation for test generation | Bride, Hadrien; Julliand, Jacques; Masson, Pierre-Alain |
| 34 | Adapting the elitism on greedy algorithm for variable strength combinatorial test cases generation | Homaid, A.A.B.A.; Alsewari, A.A.; Zamli, Kamal Z.; Alsariera, Y.A. |
| 35 | Model Learning and Test Generation Using Cover Automata | Ipate, Florentin; Stefanescu, Alin; Dinca, Ionut |